\documentclass[twocolumn,aps,prl,amsfonts,amssymb,amsmath,showpacs,english]{
revtex4}
\usepackage[T1]{fontenc}
\usepackage[latin9]{inputenc}
\usepackage{amssymb}
\usepackage{esint}
\usepackage{babel}
\usepackage{graphicx}
\begin{document}

\title{Magnetic and superconducting structures near twin boundaries in low doped Fe-pnictides}

\author{Bo Li, Jian Li, Kevin E. Bassler, and C. S. Ting}

\affiliation{Department of Physics and Texas Center for
Superconductivity, University of Houston, Houston, Texas 77204, USA}

\begin{abstract}
The effects of twin boundaries (TBs) on the complex interaction
between magnetism and superconductivity in slightly electron-doped
Ba(Ca)(FeAs)$_2$ superconductors are investigated. The spatial
distributions of the magnetic, superconducting and charge density
orders near two different types of TBs are calculated. We find that
TBs corresponding to a 90$^\circ$ lattice rotation in the a-b plane
enable magnetic domain walls to form with only a small effective
Coulomb interaction between valance electrons, and that
superconductivity is enhanced at such TBs. Contrastingly, we find
that superconductivity is suppressed at TBs corresponding to an
asymmetrical placement of As atoms with respect to the Fe atoms in
the a-b plane.
\end{abstract}


\maketitle

The recent discovery of Fe-pnictide based superconductors offers
an alternative avenue to explore the physics of high temperature
superconductors \cite{Kamihara,Ren,Chen,Cruz,Chen08}. Similar to the
cuprates, the parent compounds of the FeAs-based superconductors
also possess antiferromagntic (AF) ground states
\cite{Cruz,Chen08}. With increasing electron or hole doping, the AF
order is suppressed and superconductivity (SC) appears in both the
cuprates and the Fe-pnictides. However, different from the cuprates,
SC and a $2\times1$ collinear AF or spin-density-wave 
order can coexist in doped Ba(FeAs)$_2$
superconductors \cite{Laplace,Julien}. Because each unit cell of
these new materials
contains two inequivalent Fe ions, different organizations of the magnetic
moments of Fe ions in both normal and superconducting states can
lead to a diverse assortment of magnetic structures and unusual electronic
properties \cite{Mazin,Gorkov}.

Recently, twin boundaries (TBs) oriented $45^\circ$ from the
x(a)-axis were observed in the normal state of
Ca(Fe$_{1-x}$Co$_{x}$As)$_{2}$ \cite{Chuang}. Across these TBs, the
a-axis of the crystal rotates by $90^\circ$, and the modulation
direction of AF order that exists is rotated by $90^\circ$ as well.
That is, $90^\circ$ magnetic domain walls (DWs) are formed at the
TBs. Also, in the SC state of underdoped
Ba(Fe$_{1-x}$Co$_{x}$As)$_{2}$ with $x<0.07$, it has been found that
the diamagnetic susceptibility is increased and that the superfluid
density is enhanced on the same type of TB \cite{Kalisky}.
Consistent with these experiments, a theoretical study \cite{Huang}
found that $90^\circ$ DWs can be formed at low doping levels and
that SC is enhanced on them. However, the DWs considered in that
study formed in the absence of TBs, being induced instead by a
strong effective Coulomb interaction between valance electrons,
while in the experiments \cite{Chuang,Kalisky} the DWs were pinned
at TBs. In this letter, in order to better understand the effects
that TBs have on the development of magnetic and superconducting
order, we investigate the magnetic, SC and charge density orders
near 4 different TB configurations using the Bogoliubov-de-Gennes
(BdG) equations for very under-doped Ca(or Ba)(FeAs)$_{2}$
compounds. Among these TBs are ones involving assymetric placement
of the As atoms above and below the Fe plane. This type of TB has
not been previously studied. This study is based upon the band model
\cite{Zhang} and the phase diagram for electron doped
Ba(Fe$_{1-x}$Co$_{x}$As)$_{2}$ \cite{Zhou}. We predict that the
enhancement or suppression of SC, the location of DWs and the
electron-density distributions are largely dependent on the nature
of TBs. Our results provide a theoretical explanation for the
relationship between TBs and enhanced SC order observed in
experiments.

Consider a Hamiltonian $H=H_0+H_{SC}+H_{int}$ that describes
the energy of valance electrons.
$H_0$ is a non-interacting energy from a two-orbital tight-banding
model, the detailed form of which
can be found in Ref.~\cite{Zhang,Zhou}. The
pairing interaction energy of the electrons is
\begin{eqnarray}
\nonumber
H_{SC}=\sum_{\textbf{i}\mu\textbf{j}\nu\sigma}\Delta_{\textbf{i}\mu\textbf{j}\nu}c^{\dagger}_{\textbf{i}\mu\sigma}
c^{\dagger}_{\textbf{j}\nu\bar{\sigma}}+\textrm{H.c.}
\end{eqnarray}
where $\Delta_{\textbf{i}\mu\textbf{j}\nu}$ is the pairing parameter
between two electrons, one at position $\textbf{i}$ with the orbital
$\mu$ and the other at position $\textbf{j}$ with orbital $\nu$, and
$c^{\dagger}_{\textbf{i}\mu\sigma}$ is the creation operator of an
electron with spin $\sigma$ at position $\textbf{i}$ with orbital
$\mu$. Here $\bar{\sigma}$ denotes the opposite spin of $\sigma$.
The mean-field magnetic interaction energy is
\cite{Zhou}
\begin{eqnarray}
\nonumber
H_{int}=(U-3J_H)\sum_{\textbf{i},\mu\ne\nu,\sigma}\langle{n_{\textbf{i}\mu\sigma}}\rangle{n_{\textbf{i}\nu\sigma}}
+(U-2J_H)\\
\nonumber
\times\sum_{\textbf{i},\mu\ne\nu,\sigma\ne\bar{\sigma}}\langle{n_{\textbf{i}\mu\bar{\sigma}}}\rangle{n_{\textbf{i}\nu\sigma}}
+U\sum_{\textbf{i},\mu,\sigma\ne\bar{\sigma}}\langle{n_{\textbf{i}\mu\bar{\sigma}}}\rangle{n_{\textbf{i}\mu\sigma}}
\end{eqnarray}
where $U$ is the on-site
Coulomb interaction, $J_H$ is the Hund's coupling,
$n_{\textbf{i}\mu\sigma}$
is the electron number operator,
and
$\langle n_{\textbf{i}\mu\sigma} \rangle$
is the local electron density.
The
eigenvalues and eigenfunctions of the total Hamiltonian
$H$ can be obtained by self-consistently solving the BdG equations
\begin{eqnarray}
\nonumber \sum_{\textbf{j},\nu}\left( \begin{array}{ccc}
H_{\textbf{i}\mu\textbf{j}\nu\sigma}&\Delta_{\textbf{i}\mu\textbf{j}\nu}
\\ \Delta^{*}_{\textbf{i}\mu\textbf{j}\nu}&-H^{*}_{\textbf{i}\mu\textbf{j}\nu\bar{\sigma}}
\end{array} \right)  \left( \begin{array}{ccc}
u^{n}_{\textbf{j}\nu\sigma} \\ v^{n}_{\textbf{j}\nu\bar{\sigma}}
\end{array} \right)=E_n\left(
\begin{array}{ccc}u^{n}_{\textbf{i}\mu\sigma} \\
v^{n}_{\textbf{i}\mu\bar{\sigma}} \end{array} \right)
\end{eqnarray}
where
\begin{eqnarray}
\nonumber
H_{\textbf{i}\mu\textbf{j}\nu\sigma}=-t_{\textbf{i}\mu\textbf{j}\nu}+[U\langle
n_{\textbf{i}\mu\bar{\sigma}}\rangle +(U-2J_H)\langle
n_{\textbf{i}\bar{\mu}\bar{\sigma}}\rangle  \\
\nonumber +(U-3J_H)\langle
n_{\textbf{i}\bar{\mu}\sigma}\rangle-t_0]\delta_{\textbf{i}\textbf{j}}\delta_{\mu\nu}
\end{eqnarray}
is the matrix element
of $H$ with the same spin $\sigma$ between the orbital $\mu$ at
position $\textbf{i}$ and the orbital $\nu$ at position $\textbf{j}$,
and $t_0$ is the chemical potential. The pairing parameter
$\Delta_{\textbf{i}\mu\textbf{j}\nu}$ and the local electron densities
$\langle n_{\textbf{i}\mu\uparrow}\rangle$
and
$\langle n_{\textbf{i}\mu\downarrow}\rangle$
satisfy the following
self-consistent conditions
\begin{eqnarray}
\nonumber \Delta_{\textbf{i}\mu\textbf{j}\nu} & = &
\frac{V_{\textbf{i}\mu\textbf{j}\nu}}{4}\sum_{n}(u^{n}_{\textbf{i}\mu\uparrow}
v^{n*}_{\textbf{j}\nu\downarrow} +u^{n}_{\textbf{j}\nu\uparrow}
v^{n*}_{\textbf{i}\mu\downarrow})\tanh(\frac{E_n}{2k_BT})
\\
\nonumber \langle n_{\textbf{i}\mu\uparrow}\rangle & = &
\sum_{n}\left|u^{n}_{\textbf{i}\mu\uparrow}\right|^{2}f(E_n)
\\
\nonumber \langle n_{\textbf{i}\mu\downarrow}\rangle & = &
\sum_{n}\left|v^{n}_{\textbf{i}\mu\downarrow}\right|^{2}[1-f(E_n)]
\end{eqnarray}
where $V_{\textbf{i}\mu\textbf{j}\nu}$ is the pairing strength and
$f(x)$ is the Fermi-Dirac distribution function. The SC order
parameter at position $\textbf{i}$ is defined as $
\Delta_{\textbf{i}}\equiv\frac{1}{4}(\Delta_{\textbf{i},\textbf{i}+\hat{x}+\hat{y}}+\Delta_{\textbf{i},\textbf{i}-\hat{x}-\hat{y}}
\Delta_{\textbf{i},\textbf{i}+\hat{x}-\hat{y}}+\Delta_{\textbf{i},\textbf{i}-\hat{x}+\hat{y}})
$, the local magnetic moment at position
$\textbf{i}$ is defined
as $ m_{\textbf{i}}\equiv\frac{1}{2}\sum_{\mu}(\langle
n_{\textbf{i}\mu\uparrow}\rangle-\langle
n_{\textbf{i}\mu\downarrow}\rangle) $,
and the total charge density at position
$\textbf{i}$ is defined
as $
\langle n_{\textbf{i}}\rangle
\equiv\sum_{\mu}(\langle
n_{\textbf{i}\mu\uparrow}\rangle+\langle
n_{\textbf{i}\mu\downarrow}\rangle) $.
The chemical potential
$t_0$ is determined by the electron filling per site $n$ ($n=2+x$),
and for
the value of the hopping terms
$t_{\textbf{i}\mu\textbf{j}\nu}$ are assumed to be
$t_{1-4}=1,0.4,-2,0.04$ \cite{Zhang,Zhou}.
Only the electron pairings of the same orbital between the
next-nearest-neighbor Fe sites are considered. For example, we
choose $V_{\textbf{i}\mu\textbf{j}\nu}=1.4$ for $\mu=\nu$ and
$\left|\textbf{i}-\textbf{j}\right|=\sqrt{2}$, and zero for all
other cases. This choice of the pairing potential implies that the
SC order has $s_{\pm}$-wave symmetry \cite{Seo,Mazin08}.

The phase diagram of the electron-doped Ba(Fe$_{1-x}$Co$_x$As)$_2$
compounds as a function of temperature $T$ and doping $x$ has
been qualitatively mapped
out with $U=3.4$ and $J_H=1.3$ \cite{Zhou}. When both $T$ and
$x$ are small,
2$\times$1
collinear AF order is found, with SC order uniformly distributed over
the sample. However, this AF order is
unstable against the formation of the $90^\circ$
magnetic DWs oriented $45^\circ$ from the x-axis as the strength of
$U$ is increased to $U=4.8$ for small $x$ at $T=0$
\cite{Huang}.
In the following, the spatial profiles of magnetic, SC and charge density
order near four different types of TB at $T=0$ will be investigated. Throughout
this work, we set $x=0.04$, $U=3.8$ and $J_H=1.3$. Note that DWs do not
form spontaneously in the absence of TBs at these parameter values,
instead the superconducting and charge density orders are uniform
and AF order exists.

\emph{A. Diagonal TB of the Lattice}-
The lattice in the Fe-plane of these compounds is almost square,
having slightly different
lattice constants $a$ and $b$ along $x$- and
$y$-directions \cite{Kim}. TBs can thus be formed by exchanging the
lattice constants $a$ and $b$ on the opposite side of the TB.
Figure 1(a)
shows the structure of a single such TB oriented at $45^\circ$ with
respect to the $x$-axis.
Since there is
only a small difference in the magnitudes of $a$ and $b$,
this TB can be accounted for by assuming slightly different
nearest neighbor hoping terms
$t_a=1.0$ and $t_b=1.2$.
To analyze the effects of this type of TB,
we considered a $28\times28$ lattice with periodic boundary conditions
divided into 4 different domains separated
by three parallel TBs (not shown) along the
lines $y=x+14$, $y=x$ and $y=x-14$.
As shown in
Fig.1(b), there are three $90^\circ$ DWs formed and pinned on the
TBs. The patterns of these quantities are very similar to those
found in \cite{Huang} without the TBs (see Fig.2(a) to 2(c) in
\cite{Huang}). However, in this case, the DW forms with a smaller
value of the effective Coulomb interaction $U$,
indicating that the existence of this type of TB is beneficial to the
formation of the 90$^\circ$ DWs.
The solutions presented in Fig.1 are always stable
against the uniform $2\times1$ collinear AF order \cite{Zhou}.
\begin{figure}[t]
\includegraphics[width=1.68in] {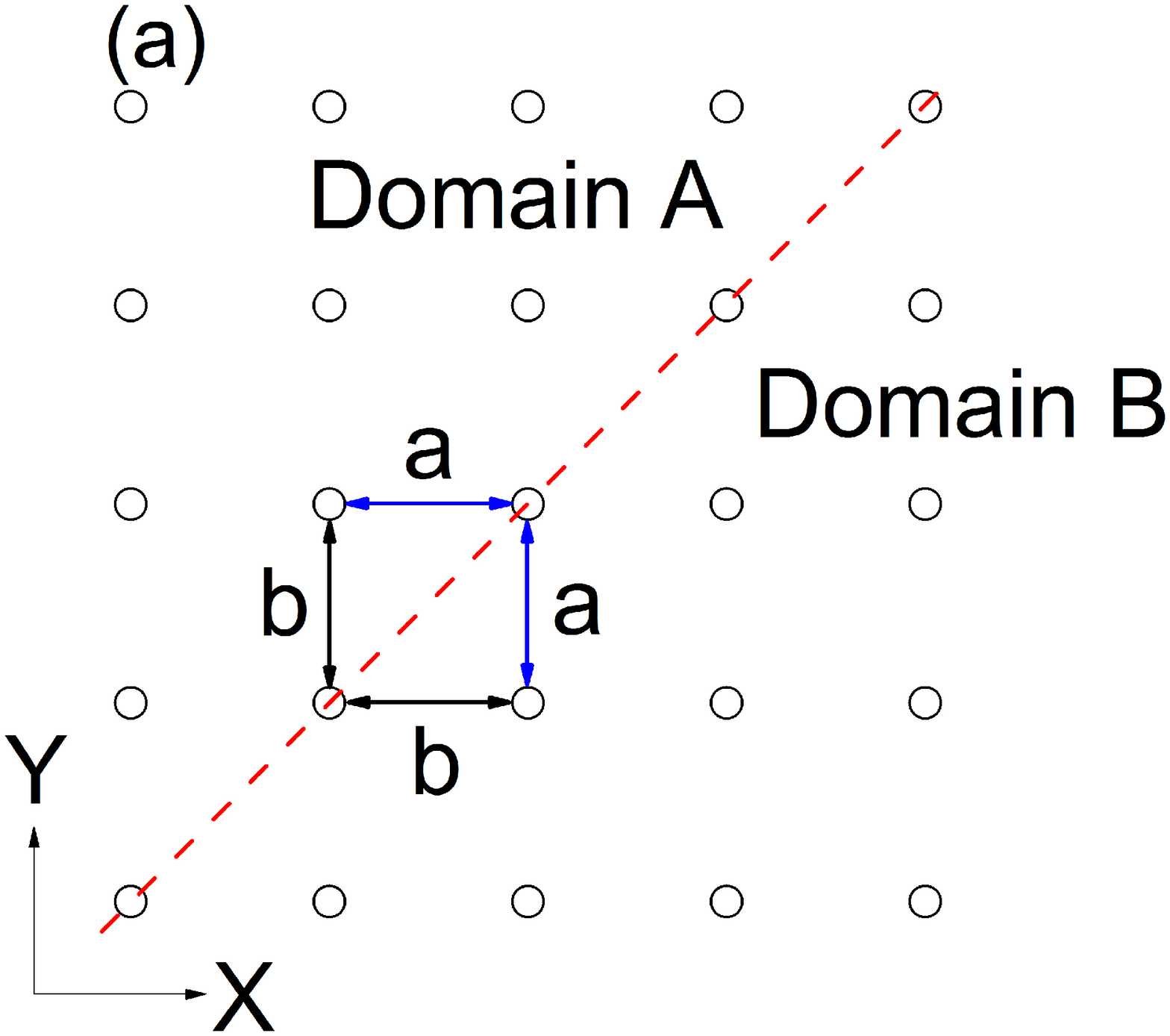}
\includegraphics[width=1.68in] {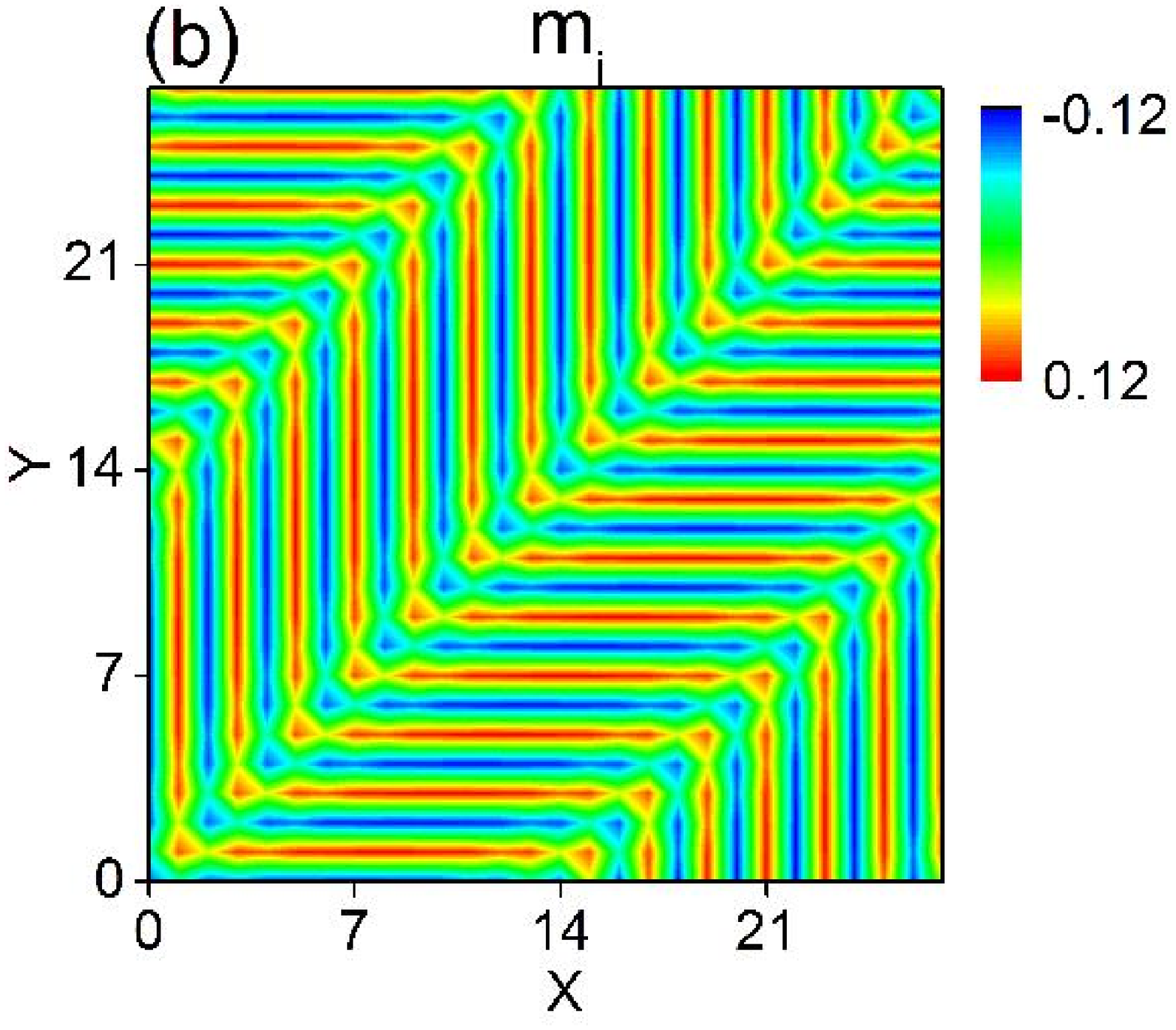}
\includegraphics[width=1.68in] {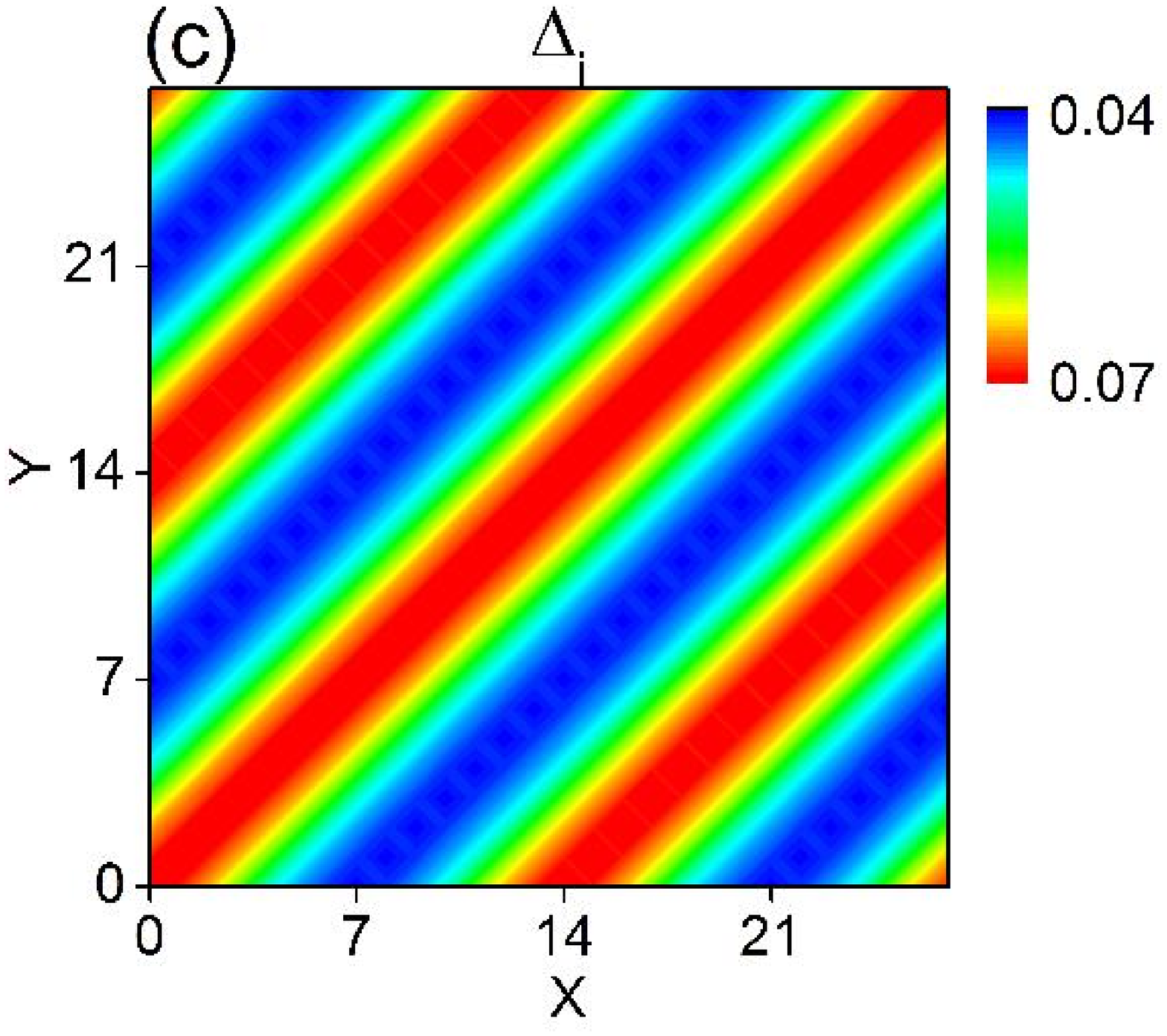}
\includegraphics[width=1.68in] {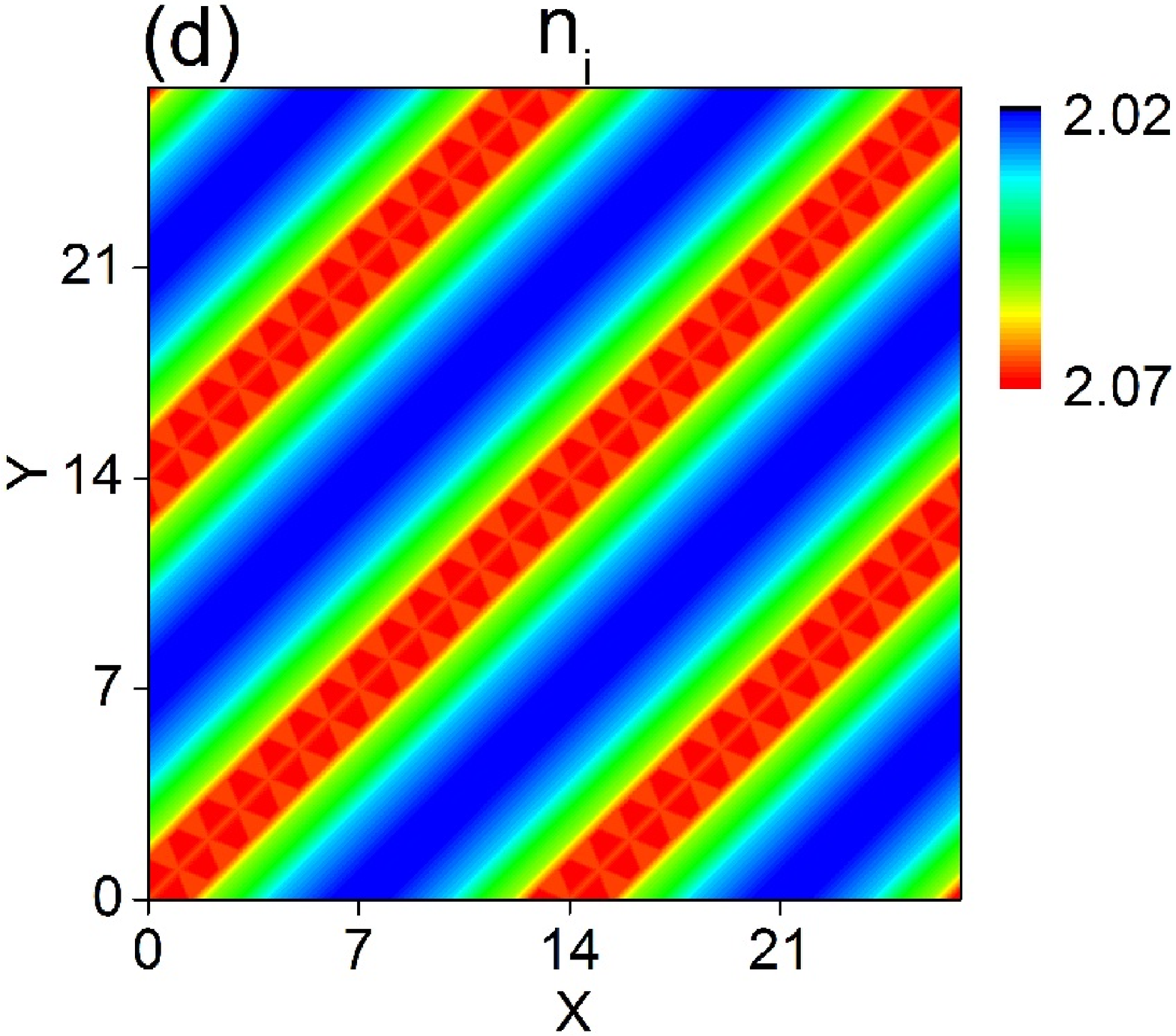}
\caption{(a) The lattice structure near a diagonal TB
(red dashed line), the open circles represent the positions of Fe
atoms, a (blue solid line) and b (black solid line) are the lattice
constants along x and y directions in domain A. Corresponding
spatial profiles of (b) the magnetic order, (c) the
superconducting order, and (d) the charge density order are
presented.}\label{Fig1}
\end{figure}

Similar to the results without TBs \cite{Huang}, the SC as well
as the charge density get significantly enhanced on the DWs,
which occur at the TBs, and
suppressed in the middle of the magnetic domains (see Fig.1(c) and
1(d)). All of this is in good agreement with experiments
\cite{Chuang,Kalisky}.
It is important to
note that the lattices on both sides of the TB should be well
matched at the TB, and that each of the unit cells along the TB is only
slightly deformed from the square shape.
Thus, we do not expect that
scattering of the electrons from any disorder due to the TB
would be strong.

\emph{B. Parallel TB of the Lattice}- A TB formed by exchanging the
$a$ and $b$ lattice constants can also be oriented parallel to the
$x$- or $y$-axis (as shown in Fig.2(a)). We studied this case by
considering a a $28\times28$ lattice with periodic boundary
conditions divided into 3 different domains separated by two TBs
(not shown) along the lines $x=7$ and $x=20$.

\begin{figure}[t]
\includegraphics[width=1.68in] {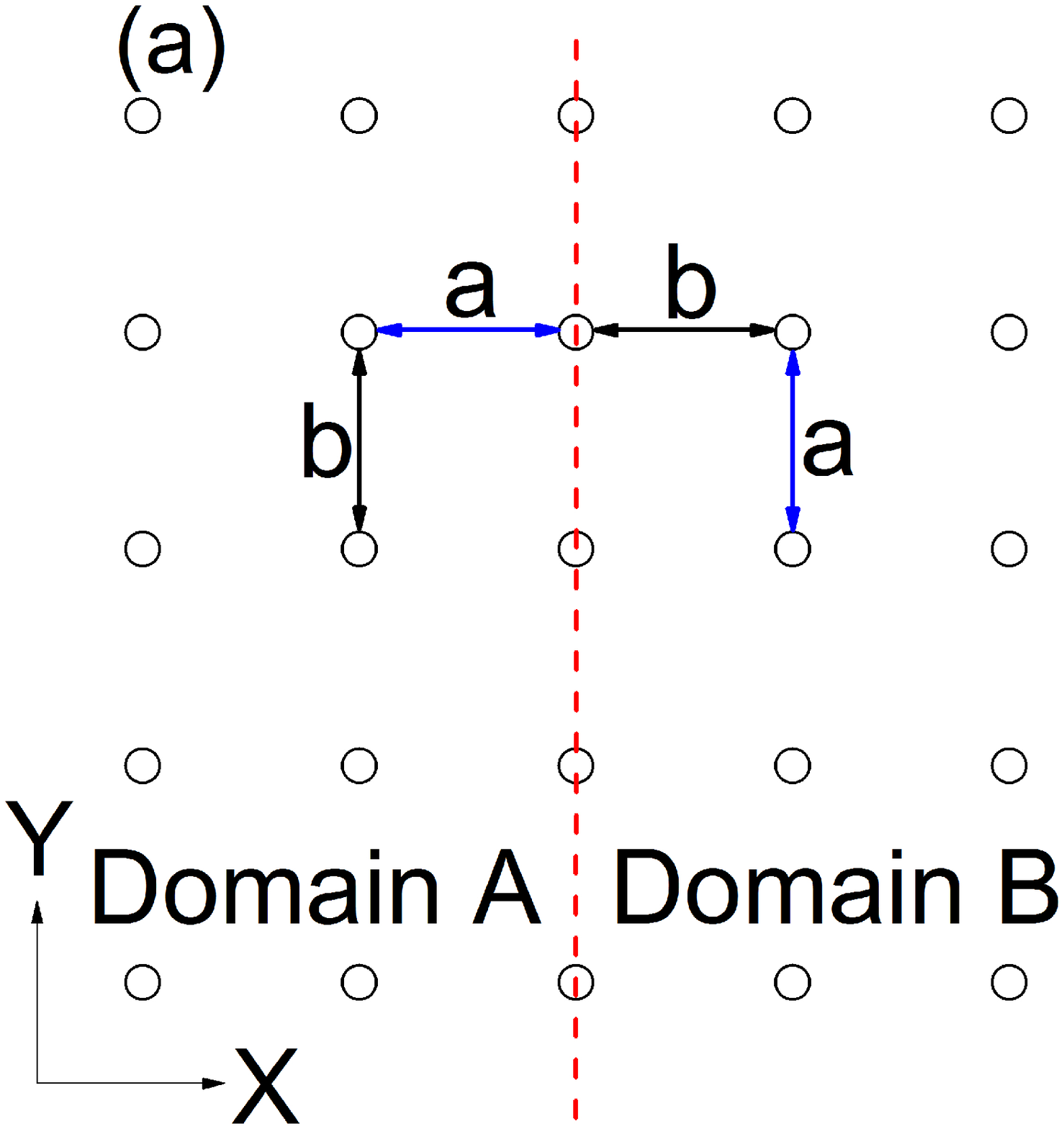}
\includegraphics[width=1.68in] {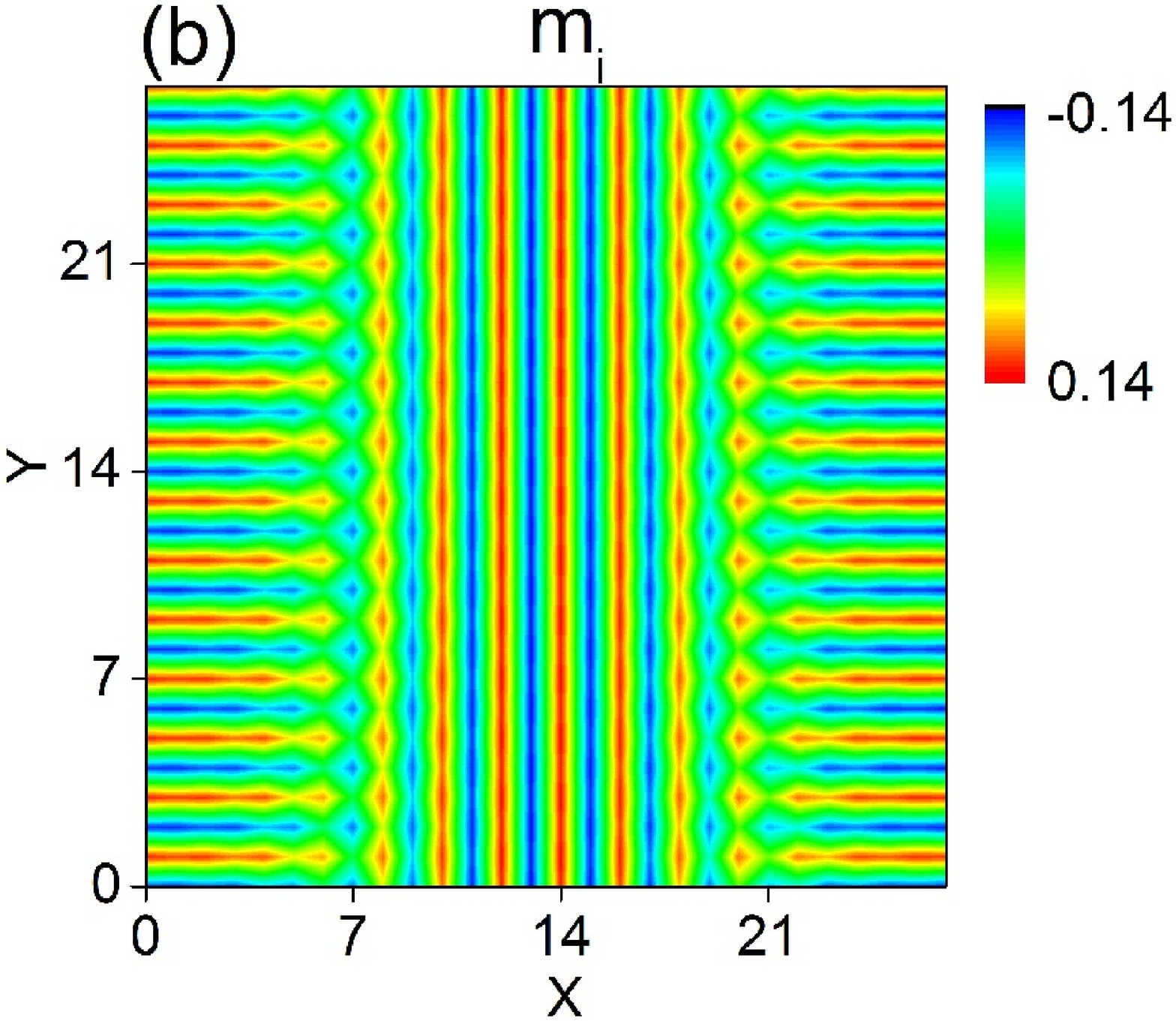}
\includegraphics[width=1.68in] {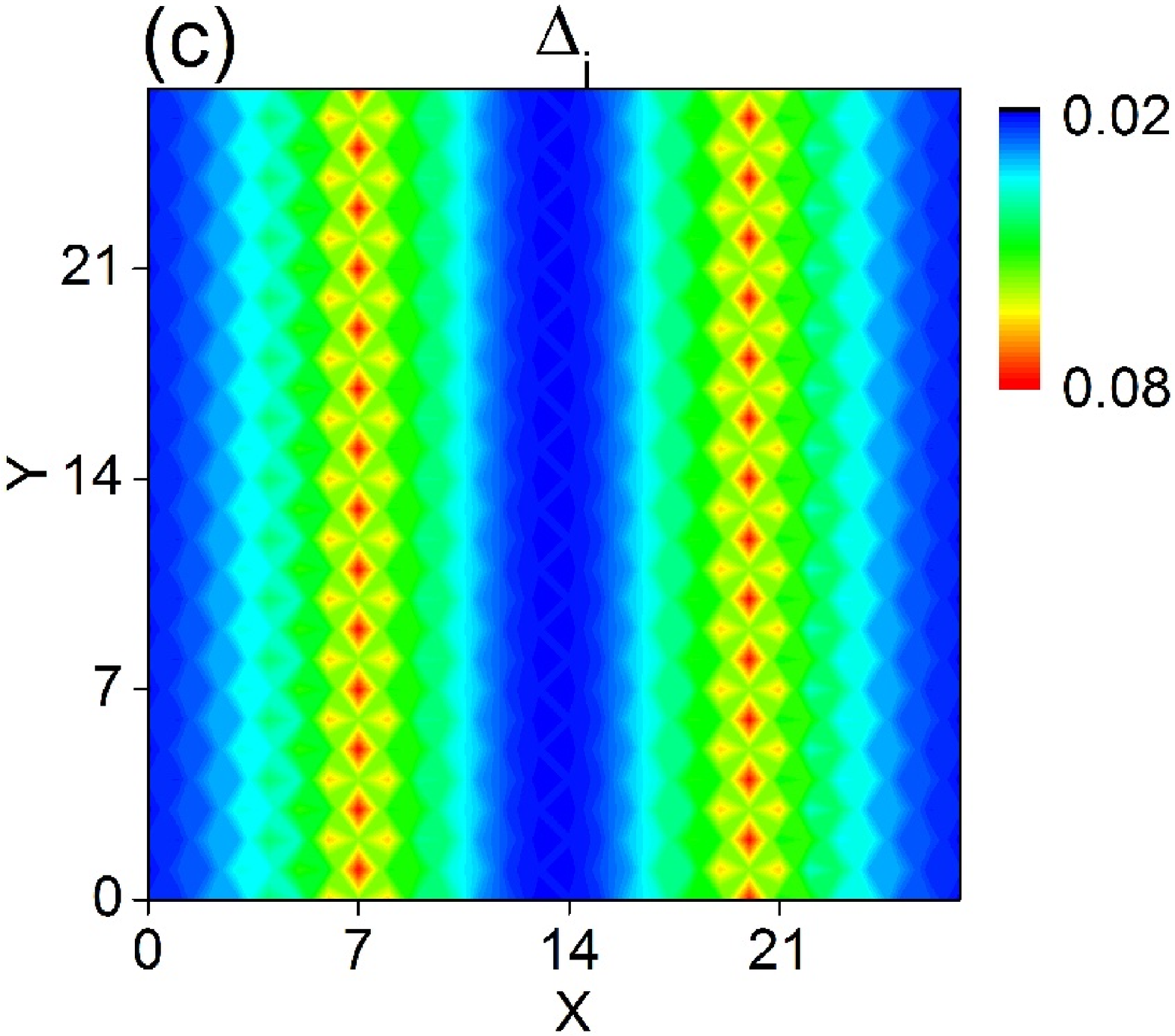}
\includegraphics[width=1.68in] {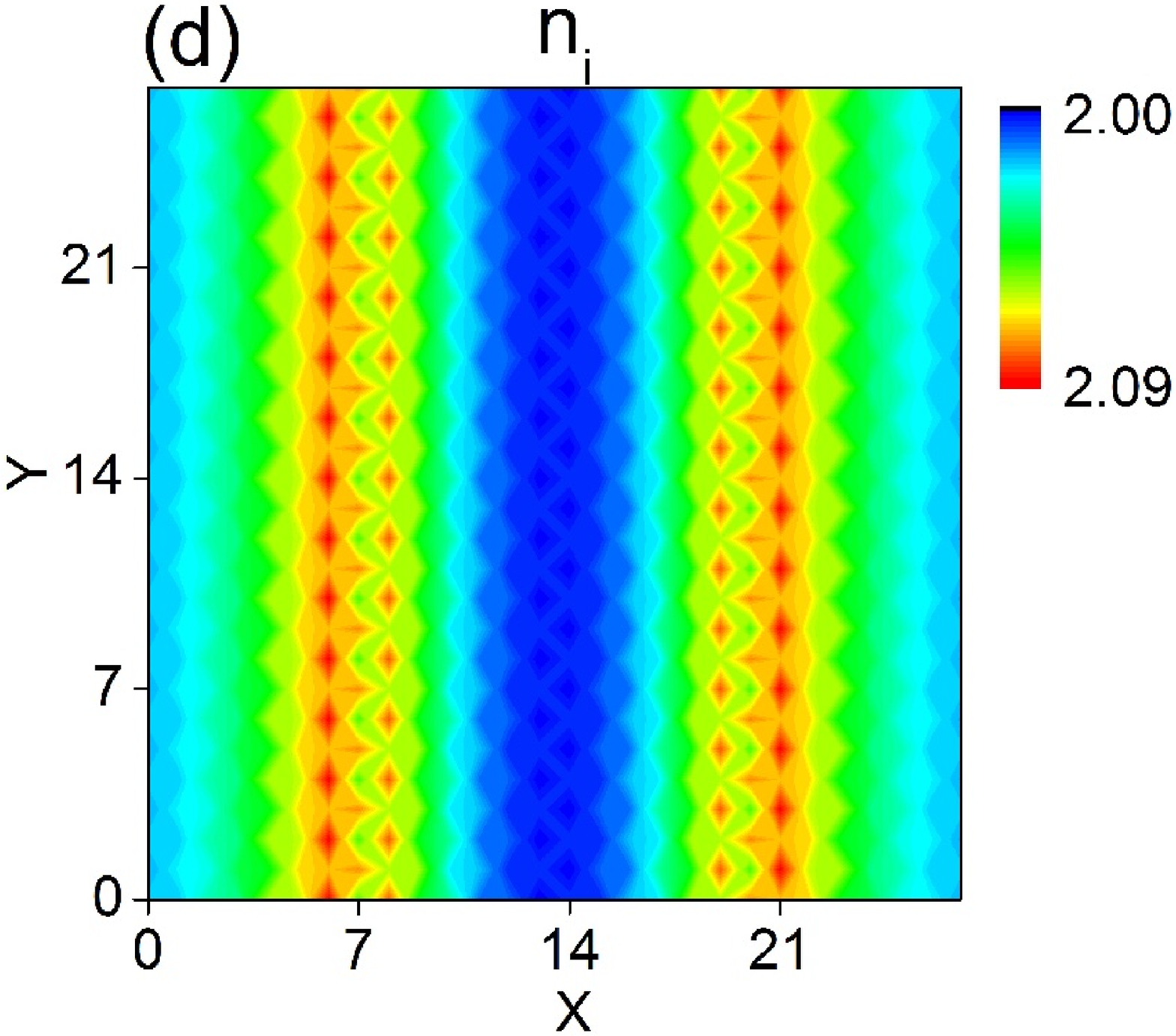}
\caption{(a) The lattice structure near a TB parallel to the $y$-direction
(red dashed
line), the open circles represent the positions of Fe atoms, a (blue
solid line) and b (black solid line) are the lattice constants along
x and y directions in domain A. Corresponding spatial profiles of
(b) the magnetic order, (c) the superconducting order, and (d) the
charge density order are presented.} \label{Fig2}
\end{figure}

Here the magnetic DWs are pinned at the TBs (see Fig.2(b)) on which
weak local ferromagnetic order appears. The SC has a periodic
modulation and is enhanced on the DWs, but suppressed in the middle
of the magnetic domains (see Fig.2(c)). A charge density wave appears
near the DWs (see Fig.2(d)) while the electron density gets
suppressed in the middle of the magnetic domains. It is important
to point out that in this case, the lattices on the opposite sides
of a TB are not well matched. Therefore, there may be considerable
scattering of the electrons due to the disorder near these TBs. If
this effect is included, we expect that the SC would get suppressed,
instead of being enhanced, on the DWs and the TBs \cite{Li}.

\begin{figure}[t]
\includegraphics[width=1.68in] {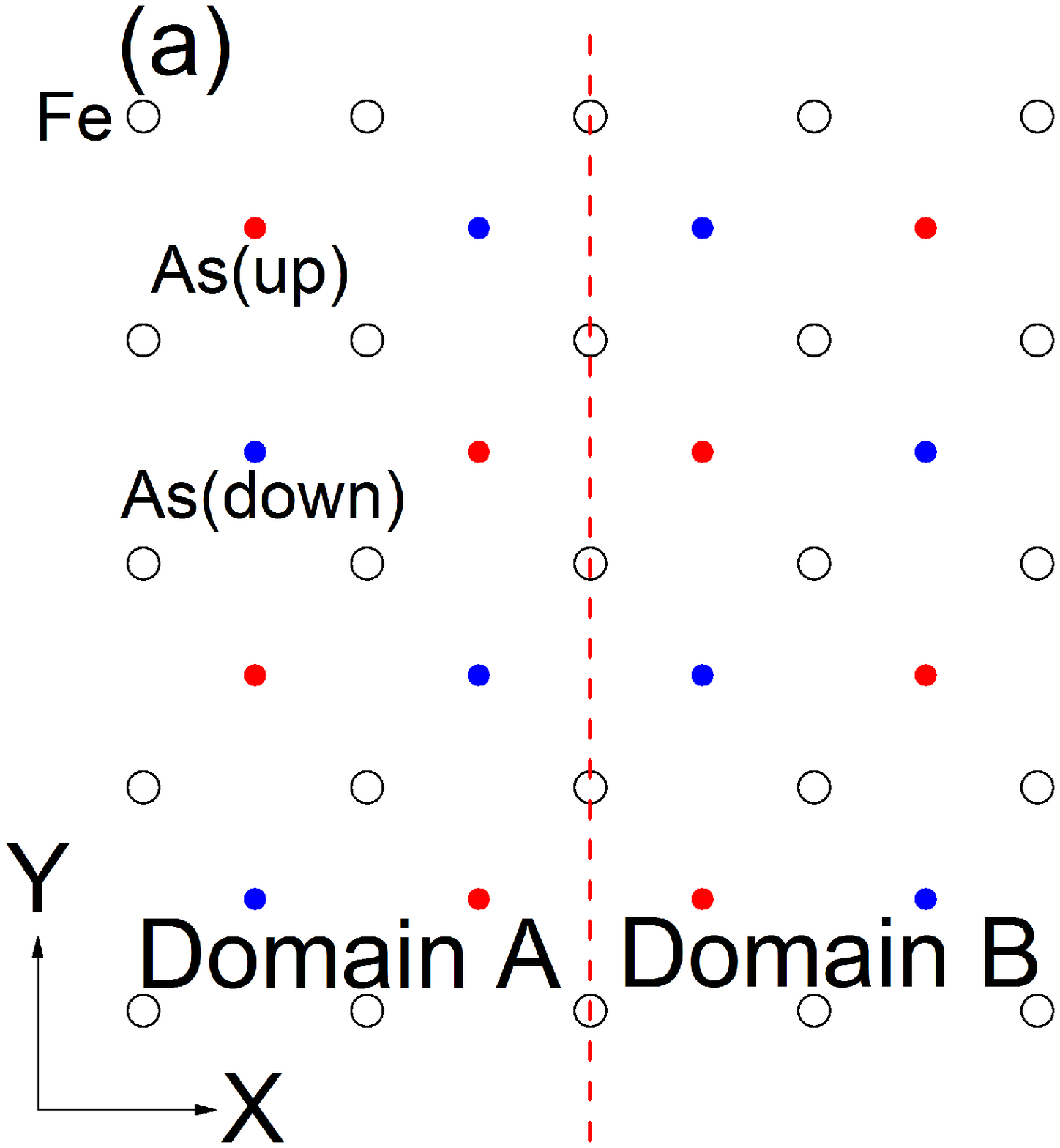}
\includegraphics[width=1.68in] {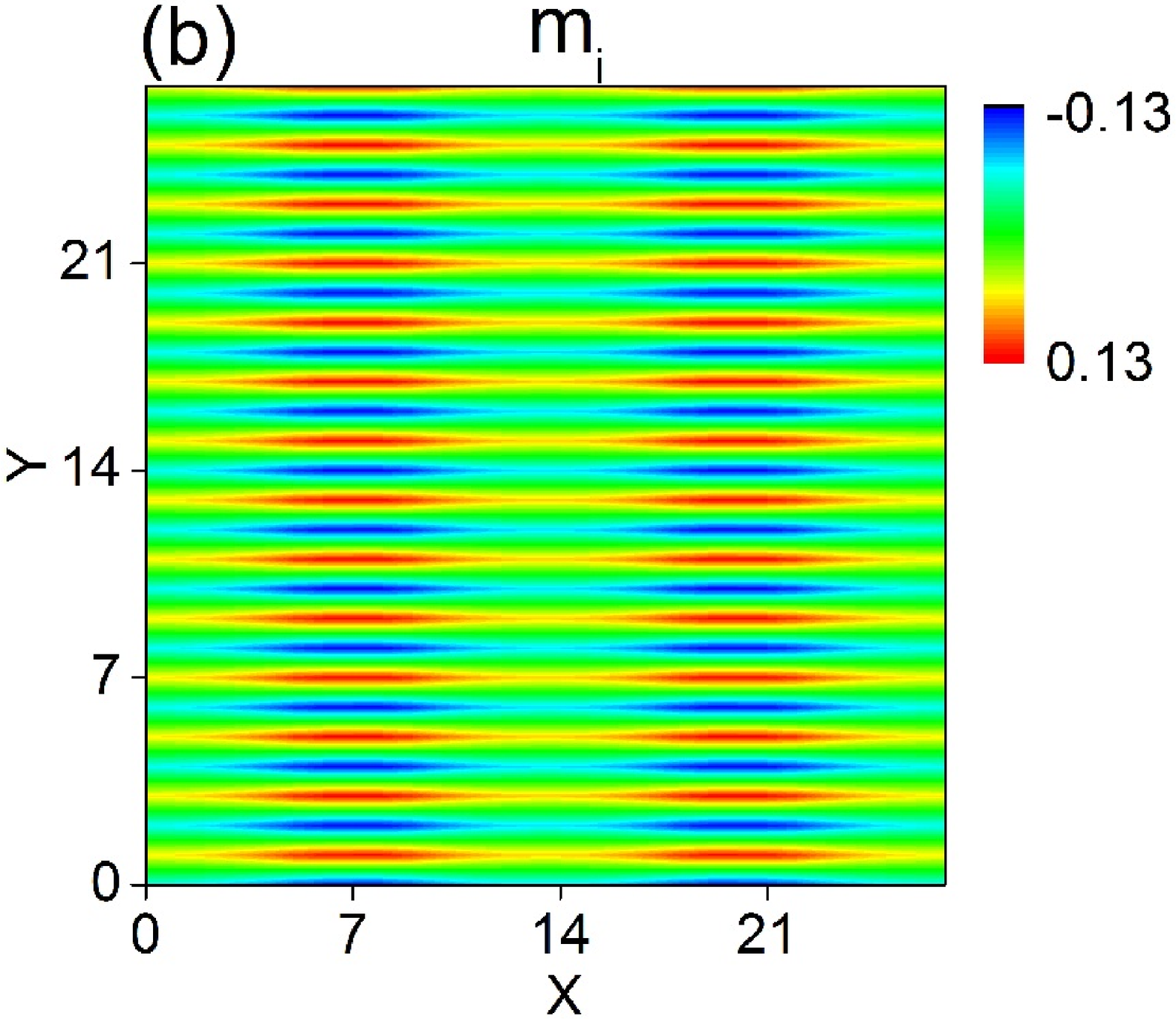}
\includegraphics[width=1.68in] {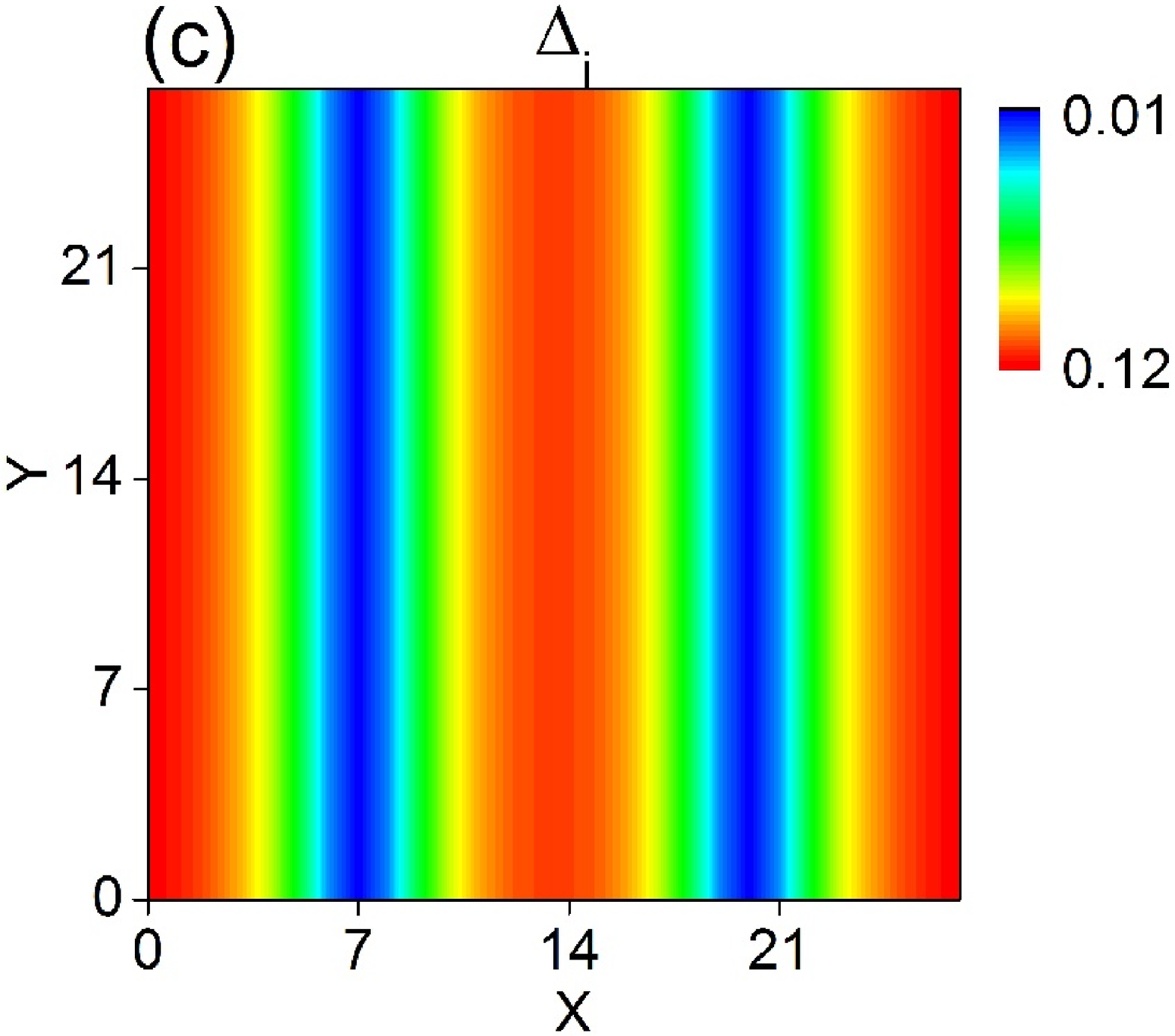}
\includegraphics[width=1.68in] {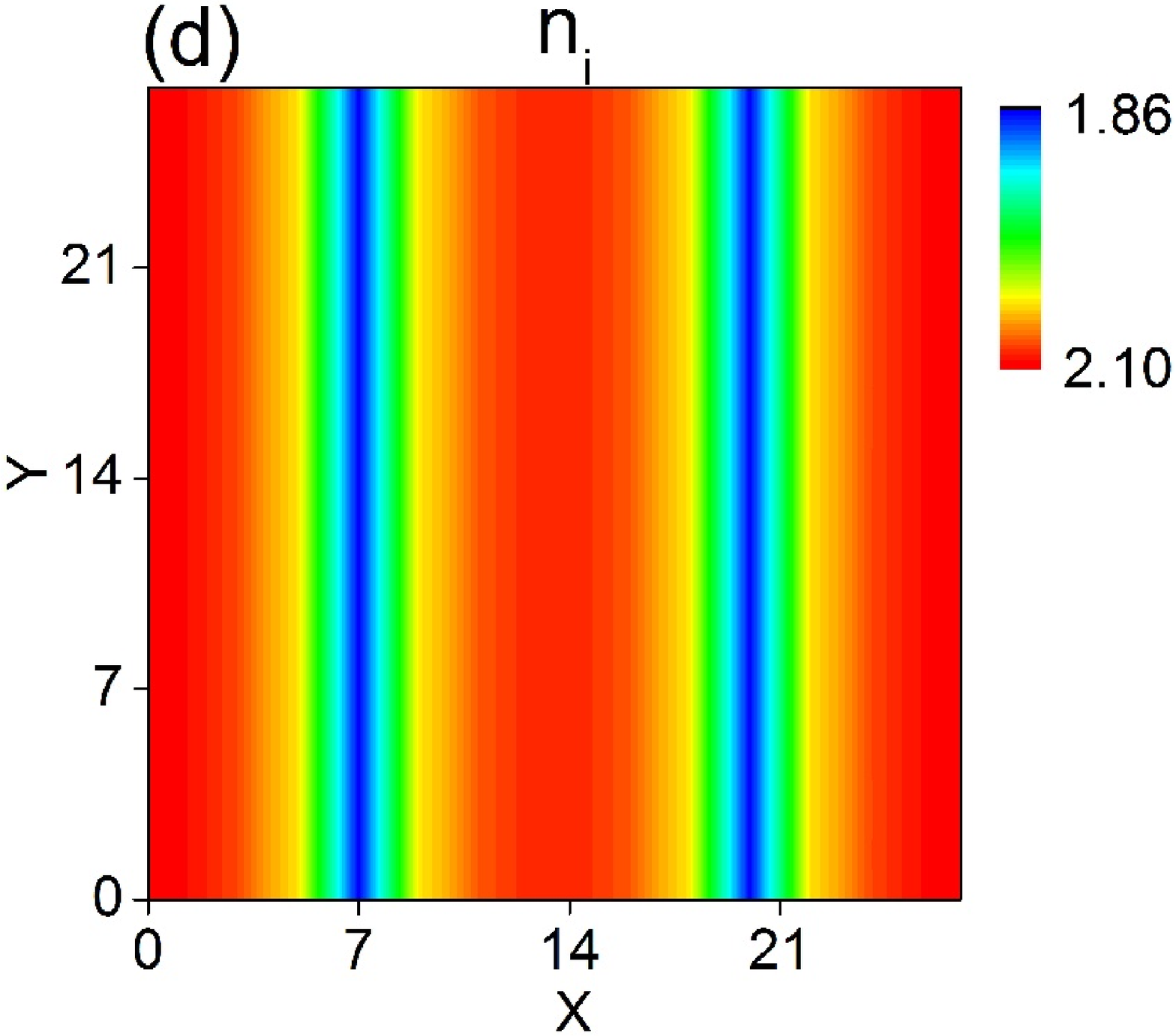}
\caption{(a) The lattice structure near a twin boundary (red dashed
line) parallel to the $y$-direction formed by misplacing As atoms.
The open circles
represent the positions of Fe atoms, and the red and blue dots respectively
denote the
As(up) and As(down) atoms. Corresponding spatial
profiles of (b) the magnetic order, (c) the superconducting order,
and (d) the charge density order are presented.} \label{Fig3}
\end{figure}

\emph{C. Parallel twin boundary due to the asymmetry of As atoms}-
Another possible TB can be generated by slipping the lattice on the
right side of the TB by a lattice constant along the y-direction
with respect on the lattice on the left of the TB. There are two
different types of As atoms in our model, we label them as As(up)
and As(down) atoms relative to the Fe plane. This can be clearly
seen from Fig.3(a), in which the TB is represented by the red-dashed
line. The crystal lattice for the FeAs layer has D$_{2d}$ symmetry,
namely the 4 nearest neighboring As atoms of a "down" As atom should
be all "up". The hopping terms between the next-nearest-neighboring
Fe ions via the hybridization of the 4p orbital with the As atom in
the middle should have different values depending on whether the As
atom is above ($t_2$) or below ($t_3$) the Fe plane \cite{Zhang,Hu}.
The $D_{2d}$ symmetry is broken by the presence of the TB. We
considered a $28\times28$ lattice with periodic boundary conditions
divided into three domains by two TBs located at $x=7$ and $x=20$
(not shown). The lattice constants are both $a$ along x- and y-axis.
Figure 3(b) shows the magnetic order is enhanced near the TBs.
Defining the magnetic DWs to be where the magnetic order is
suppressed, that is, near the middle between two TBs, then, clearly,
the DWs are not located at the TBs. On the opposite sides of a DW,
there is no change in the magnetic phase, thus we could label the
DWs as the 0$^\circ$ DWs. Figure3(c) shows that the SC is enhanced
along the DWs, and that it is suppressed near the TBs. Figure 3(d)
shows that the electron density is depleted near the TBs.
Apparently, the depleted electron density leads to strong magnetic
order that suppresses the SC order. On the DWs, the
electron density
appears to be close
to optimal doping and thus SC gets enhanced.

\emph{D. Diagonal twin boundary due to the asymmetry of
As atoms}- A TB due to missing one line of the lattice contains of
both Fe and As(down) atoms oriented along $45^\circ$ from the x-axis
is shown in Fig.4(a). The $D_{2d}$ symmetry of the lattice
is also broken by the
presence of this TB. Note that the geometry of
this TB is fundamentally different from the one showed in Fig.3(a)
since the TB does not pass through any of the Fe or As atoms.
To study this case, we considered a 30$\times$30
lattice with periodic boundary conditions and
three identical TBs oriented $45^\circ$ from the x-axis
(not shown).
The TBs are located along $y=x-15$, $y=x$ and
$y=x+15$.

\begin{figure}[t]
\includegraphics[width=1.68in] {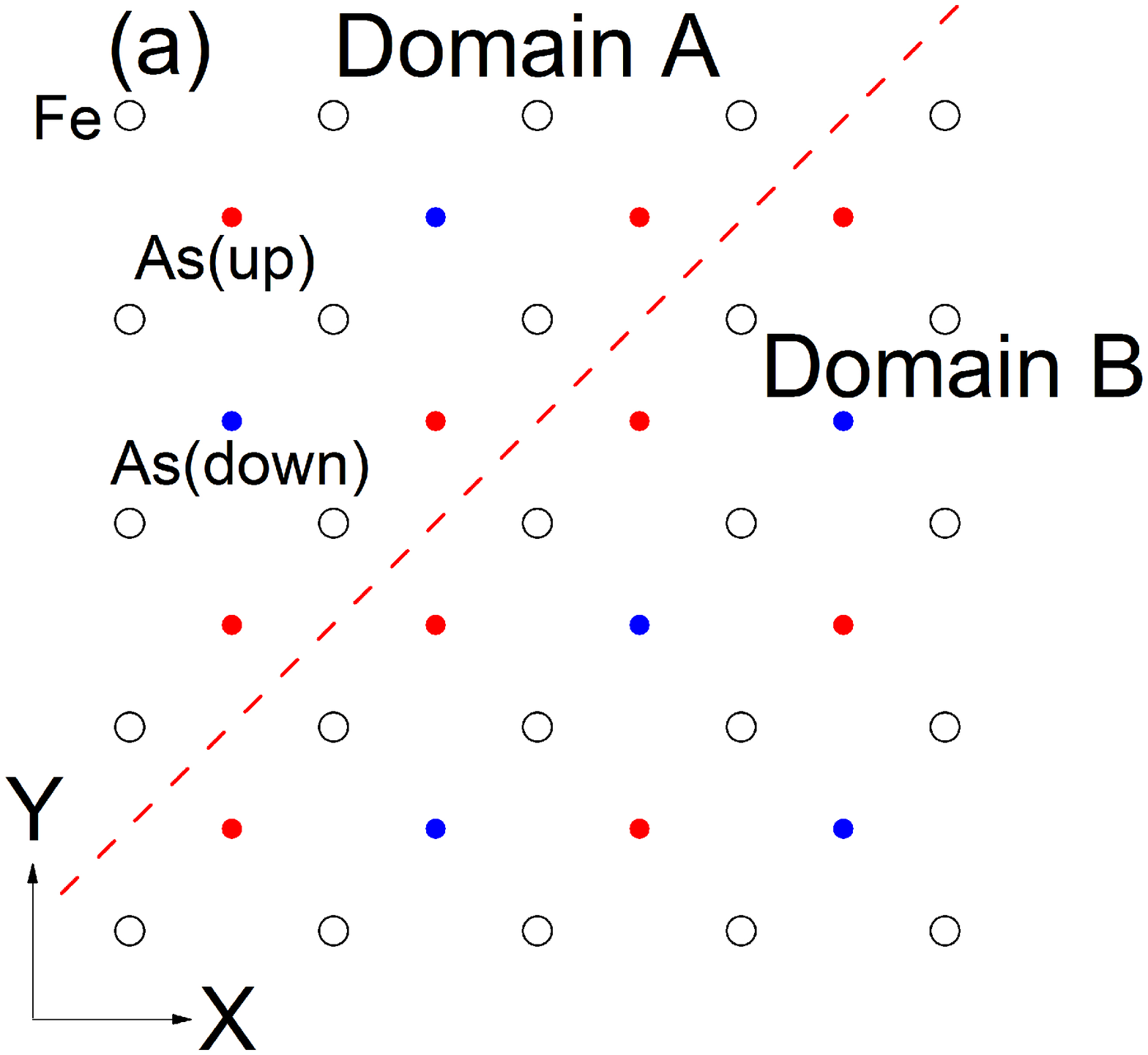}
\includegraphics[width=1.68in] {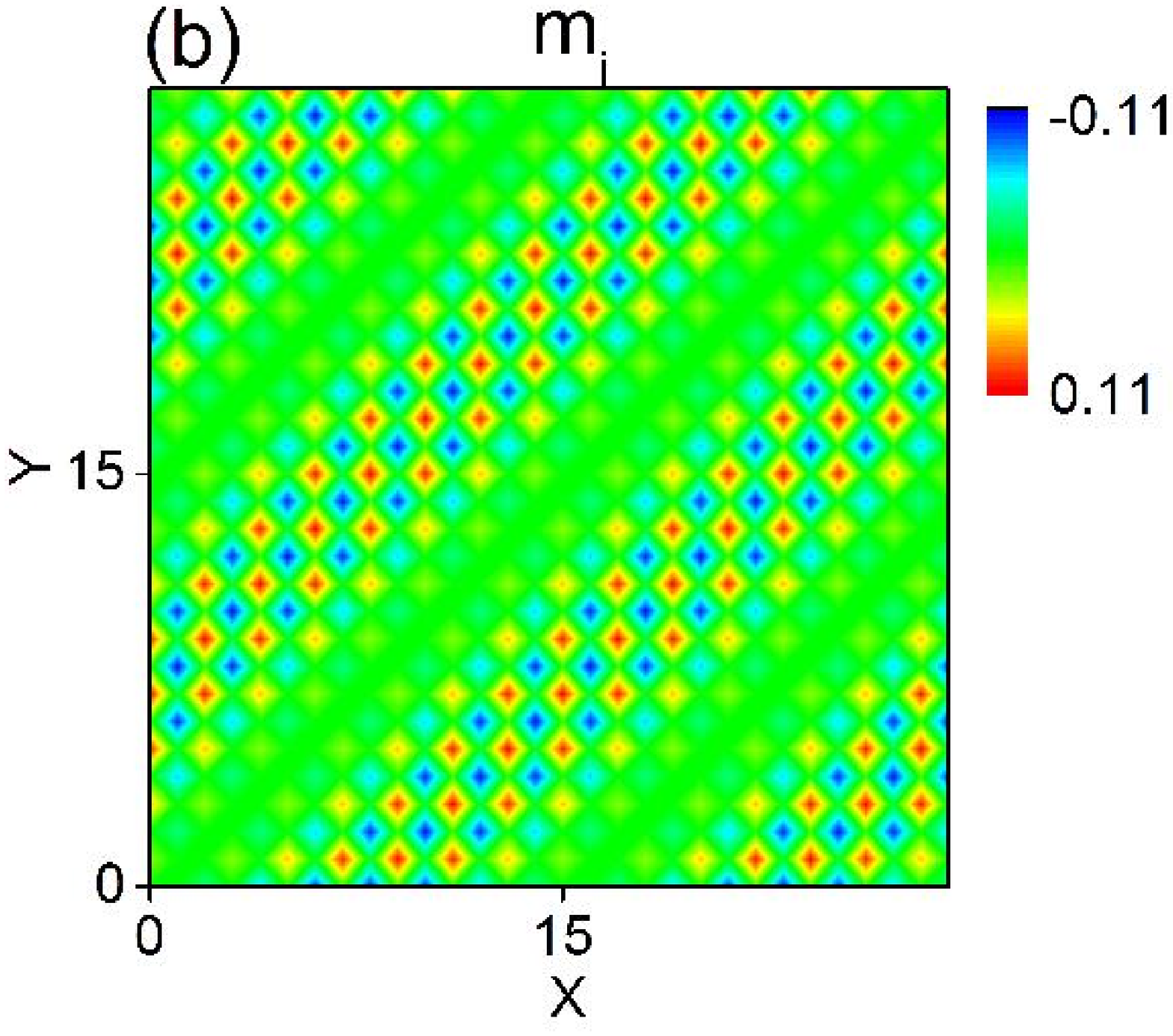}
\includegraphics[width=1.68in] {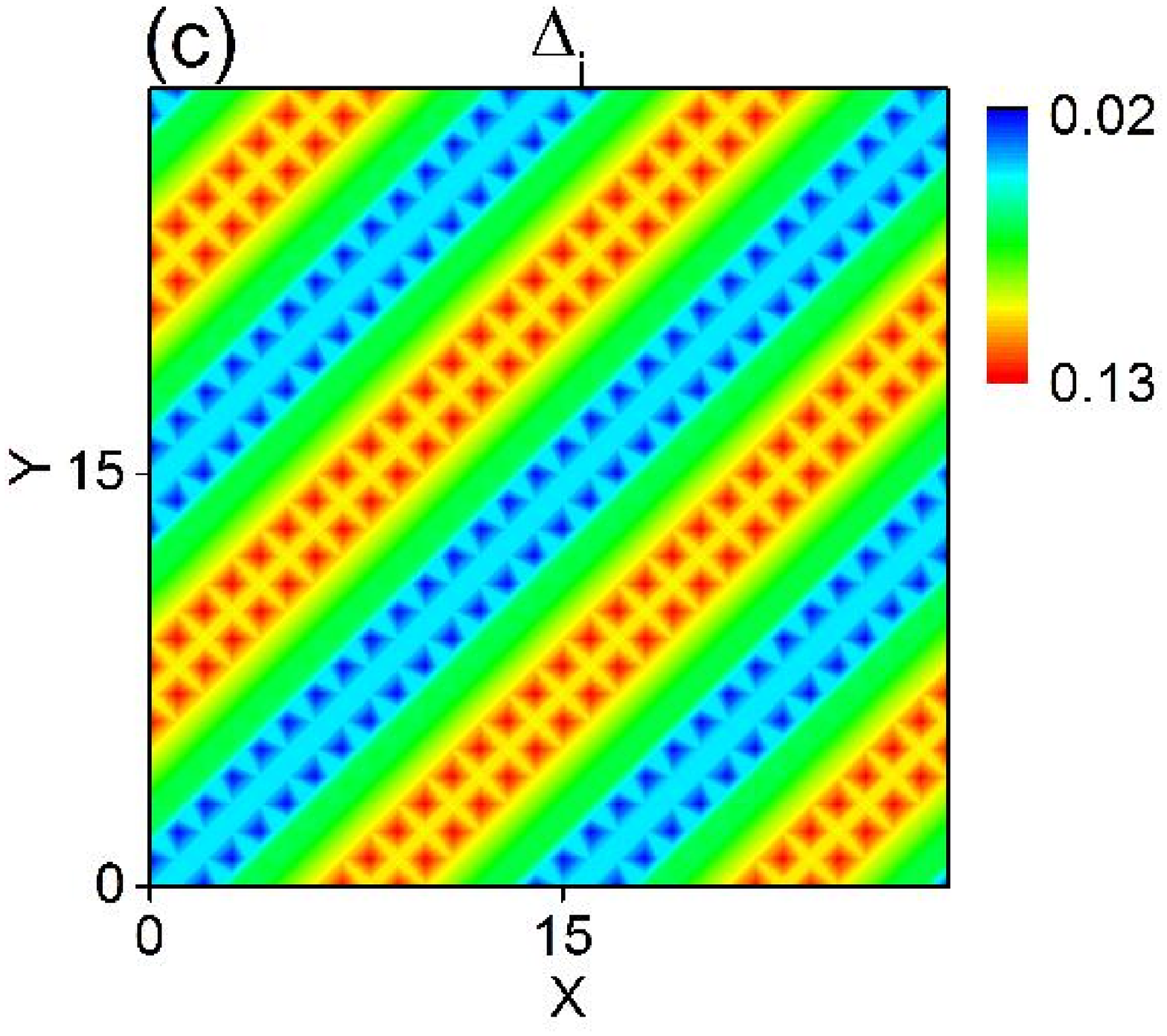}
\includegraphics[width=1.68in] {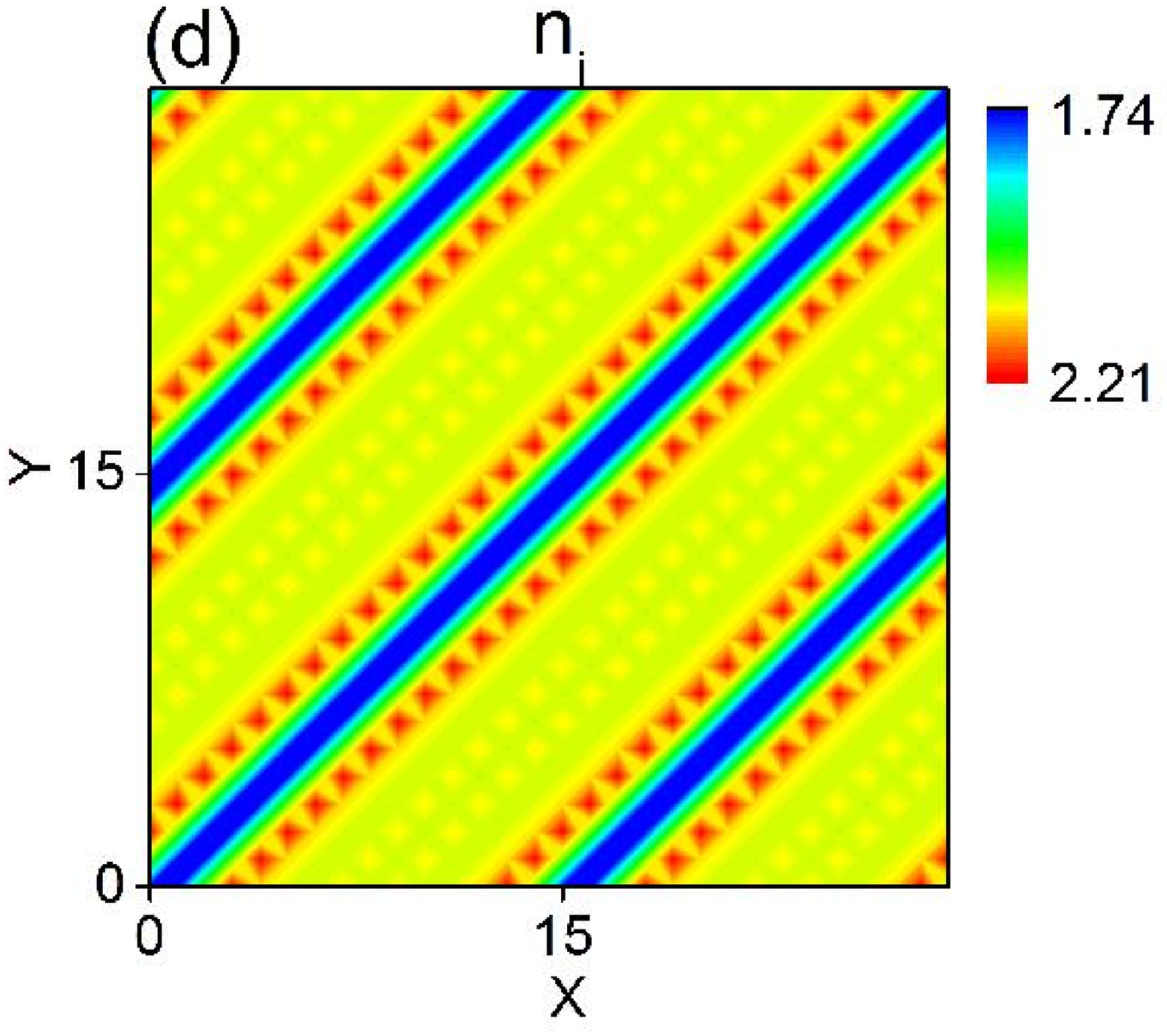}
\caption{(a) The lattice structure near a twin boundary (red dashed
line) with misplacing the As atoms along diagonal (or $45^\circ$)
direction. The open circles represent the positions of Fe atoms,
and the red and blue dots respectively denote the As(up) and As(down) atoms.
Corresponding spatial profiles of (b) the magnetic order, (c) the
superconducting order, and (d) the charge density order are presented.}
\label{Fig4}
\end{figure}

Different from the case in Fig.3(b), the magnetic order shown in
Fig.4(b) is suppressed along the TBs where DWs are located. The
magnetic domain between the TBs still has the usual $2\times1$
collinear AF structure, except  the magnetic moments are strongly
and periodically modulated along the x-axis, which may be due to
finite size effects of  the TB and the distance between two nearest
neighboring TBs. It also appears that the local $2\times1$ collinear
AF structure is replaced by a stripe-like $\sqrt{2}\times\sqrt{2}$
AF structure oriented $45^\circ$ from the x-axis. Furthermore, the
SC is enhanced in these regions (see Fig.4(c)). From Fig.4(d), note
that on the TBs or DWs the carrier density is corresponding to that
in the overly hole-doped case ($x\approx-0.3$), which explains why
the magnetic and SC orders are both suppressed on both sides of the
TB. Interestingly, stripe-like charge density waves oriented
$45^\circ$ from the x-axis occur on both sides of each TB.

Our work has considered the effects of TBs
on the complex interaction of magnetism and superconductivity in Fe-pnictides.
There are three points that need to be emphasized here. First, the
formation and the location of the DWs are strongly related to the
nature of TBs. For the four kinds of TBs studied here, the DWs in
cases A, B and D are found to be pinned at the TBs, while in case C
the DWs are separated from the TBs. Secondly, the formation of the DWs
implies that the magnetism is inhomogeneous.
This inhomogeneity strongly
affects the SC and the electron density distribution. In cases of A,
B and C, the SC is enhanced in the regions where $m_{\textbf{i}}$ is
suppressed and $\langle n_{\textbf{i}} \rangle$ is enhanced. The reasons for the
enhancement of SC are (i) the competition between the SC and
magnetism, and (ii) the
$\langle n_{\textbf{i}} \rangle$
in the enhanced regions is
close to the optimal doping level. However in case D, on the DWs
where the $m_{\textbf{i}}$ is suppressed, the electron density is
close to over(hole)-doping level, which is unfavorable to SC. As a
result, SC coexists with the magnetism in the middle of magnetic
domains. Finally, we point out that our results on the diagonal TB
of the lattice
(see Fig.1) are in good agreement with experiments
\cite{Chuang,Kalisky}. In a very recent scanning tunneling
microscopy (STM) experiment for FeSe \cite{Song}, similar types of
the TBs were detected but the SC was suppressed on the TBs. This may
be caused by the differences in lattice structures and Fermi surface
topologies of the two materials. We \cite{Li} also find that the
Coulomb-interaction-induced anti-phase DWs, as predicted in
\cite{Mazin} and measured indirectly in nuclear magnetic resonance
(NMR) experiments \cite{Xiao}, always have higher ground energies
than those of the 2x1 collinear AF order and the 90$^\circ$ DW
structures. These anti-phase DWs could form metastable states and
may occur under certain conditions. In the present work
we assume the lattices are well matched at the TBs. In the case that
the lattice on both sides of the TB are not well matched, the SC
could be suppressed by disorder scattering along the TBs. This issue
deserves further investigation.

\begin{acknowledgements}
\emph{Acknowledgement}- This work was supported by the Texas Center
for Superconductivity at the University of Houston, by the Robert
A. Welch Foundation under Grant No. E-1146, and by the NSF through grants
DMR-0908286 and DMR-1206839.
\end{acknowledgements}

\end{document}